\documentclass[aps,pra,twocolumn,superscriptaddress,nofootinbib]{revtex4-1}

\usepackage{graphicx}
\usepackage{dcolumn}
\usepackage{bm}
\usepackage{amsmath}
\usepackage{amssymb}
\usepackage{latexsym}
\usepackage{epsfig}
\usepackage{amsbsy}
\usepackage{array}
\usepackage{amssymb}
\usepackage{setspace}
\usepackage{bm}
\usepackage{epstopdf}
\usepackage{subfigure}

\def\sint{\ifmmode{- \!\!\!\!\!\! \int}
    \else{\hbox{$- \!\!\!\! \int \ $}}\fi}

\begin{document}

\title{Noise robustness of synchronization of two nanomechanical resonators \\ coupled to the same cavity field}

\author{Wenlin Li}
\affiliation{School of Science and Technology, Physics Division, University of Camerino, I-62032 Camerino (MC), Italy}
\author{Paolo~Piergentili}
\affiliation{School of Science and Technology, Physics Division, University of Camerino, I-62032 Camerino (MC), Italy}
\affiliation{INFN, Sezione di Perugia, I-06123 Perugia, Italy}
\author{Jie Li}
\affiliation{School of Science and Technology, Physics Division, University of Camerino, I-62032 Camerino (MC), Italy}
\affiliation{Zhejiang Province Key Laboratory of Quantum Technology and Device,
Department of Physics and State Key Laboratory of Modern Optical Instrumentation, Zhejiang University, Hangzhou, Zhejiang, 310027, China}
\author{Stefano~Zippilli}
\affiliation{School of Science and Technology, Physics Division, University of Camerino, I-62032 Camerino (MC), Italy}
\author{Riccardo~Natali}
\affiliation{School of Science and Technology, Physics Division, University of Camerino, I-62032 Camerino (MC), Italy}
\affiliation{INFN, Sezione di Perugia, I-06123 Perugia, Italy}
\author{Nicola~Malossi}
\affiliation{School of Science and Technology, Physics Division, University of Camerino, I-62032 Camerino (MC), Italy}
\affiliation{INFN, Sezione di Perugia, I-06123 Perugia, Italy}
\author{Giovanni~Di~Giuseppe}
\affiliation{School of Science and Technology, Physics Division, University of Camerino, I-62032 Camerino (MC), Italy}
\affiliation{INFN, Sezione di Perugia, I-06123 Perugia, Italy}
\author{David~Vitali}
\affiliation{School of Science and Technology, Physics Division, University of Camerino, I-62032 Camerino (MC), Italy}
\affiliation{INFN, Sezione di Perugia, I-06123 Perugia, Italy}
\affiliation{CNR-INO, L.go Enrico Fermi 6, I-50125 Firenze, Italy}

\date{\today}
\begin{abstract}
We study synchronization of a room temperature optomechanical system formed by two resonators coupled via radiation pressure to the same driven optical cavity mode. By using stochastic Langevin equations and effective slowly-varying amplitude equations, we explore the long-time dynamics of the system. We see that thermal noise can induce significant non-Gaussian dynamical properties, including the coexistence of multi-stable synchronized limit cycles and phase diffusion. Synchronization in this optomechanical system is very robust with respect to thermal noise: in fact, even though each oscillator phase progressively diffuses over the whole limit cycle, their phase difference is locked, and such a phase correlation remains strong in the presence of thermal noise.
\end{abstract}
\pacs{75.80.+q, 77.65.-j}
\maketitle
\section{Introduction}  
Spontaneous synchronization of two oscillators induced by a weak mutual interaction has been investigated extensively since its first observation by Huygens in the late
$1600$s~\cite{C. Huygens}. In the last decade, research in this field has been gradually extended into the micro and nano domain, where quantum effects may manifest themselves.
Some representative theories from classical synchronization, such as the analysis based on the Kuramoto model~\cite{Kuramoto1984,Acebron2005}, carry over to the mean-field dynamics of quantum systems~\cite{Heinrich2011,Holmes2012,Lee2014,Witthaut2017}. On this basis, synchronization phenomena have been predicted theoretically or observed experimentally in various microscopic systems, such as van der Pol (VdP) oscillators~\cite{Lee2014,Witthaut2017,Lee2013,Weiss2017,Jessop2019}, atomic ensembles~\cite{Xu2014,Hush2015,Stefanatos2019}, cavity/circuit electrodynamics systems~\cite{Nigg2018,Cardenas2019} and optomechanical systems (OMSs)~\cite{Heinrich2011,Mari2013,Ludwig2013,Bagheri2013,Zhang2015,Ying2014,Weiss2016,Li2016,Bemani2017}. On the other hand, quantum effects may be responsible for some differentiation between classical and quantum synchronization. Some approaches have been developed to address this problem, by introducing fluctuations and the constraints imposed by the Heisenberg uncertainty principle into their quantitative analysis~\cite{Manzano2013,Mari2013,Li2017}. Subsequently, the relation between synchronization and quantum correlations, such as entanglement and discord, have been explored in Refs.~\cite{Giorgi2012,Giorgi2013,Mari2013,Bemani2017,Ameri2015,Roulet2018,Stefanatos2017,Bergholm2019}, and moreover synchronization-induced quantum phase transitions have been analyzed recently in quantum many-body systems~\cite{Jin2013,Pizzi2019} and time crystals~\cite{Richerme2017}.

OMSs represent a well-developed platform to explore synchronization, with unique advantages. The radiation pressure interaction between optical and mechanical modes can induce a variety of nonlinear behaviors by only adjusting the corresponding pump laser~\cite{Marquardt2006,Bakemeier2015}. In particular, the mechanical oscillators can be driven into limit cycles, a prerequisite for exploring synchronization~\cite{Roulet2018,Kwek2018}, when driving with a blue-detuned~\cite{Mari2013,Marquardt2006} or gently modulated~\cite{Mari2009} pump laser. Moreover in OMSs one can measure with great sensitivity both position and momentum of the mechanical oscillator~\cite{Aspelmeyer2014,Bawaj2015,Oconnell2010}. Synchronization in OMSs has been investigated up to now in a variety of multimode structures~\cite{Mari2013,Ying2014,Cabot2017,Bemani2017,Zhang2015}, and a common scheme is based on coupling several mechanical modes to a common optical mode~\cite{Holmes2012,Bagheri2013,Li2016,Bemani2017,Liao2019}. Very recently, experiments have successfully coupled two membranes to a Fabry-P\'erot cavity~\cite{Piergentili2018,Gartner2018,Wei2019,Naesby2019} with enhanced optomechanical coupling due to the collective interactions~\cite{Xuereb2012,LiJ2016}, and this prompts us to investigate further the synchronization induced by the indirect coupling mediated by the cavity mode. In fact, a systematic study of the effect of noise on synchronization in OMSs is missing, because most of the studies focused onto the noiseless case only~\cite{Heinrich2011,Holmes2012}, or limited themselves to the use of mean-field approximations with linearized fluctuation terms, where all non-Gaussian properties are ignored~\cite{Mari2013,Ying2014,Li2016,Li2017,Bemani2017,Cabot2017,Liao2019}. However, recent studies of VdP oscillators and single-mode OMSs have shown that in a limit cycle, the oscillator state will deviate from the Gaussian form because of the inevitable phase diffusion~\cite{Lee2013,Navarrete-Benlloch2017,Navarrete-Benlloch2008,Rodrigues2010,kato2019}. It has been pointed out that for a single limit cycle, non-Gaussian properties induced by quantum noise occurs in the ``quantum regime'' ($g/\kappa\geq1$, where $g$ is the optomechanical coupling and $\kappa$ is the cavity decay rate)~\cite{Qian2012,Lorch2014,Ludwig2013}, even though recently it has been shown that in the presence of non-negligible thermal noise, non-Gaussian effects can occur even in the semi-classical limit ($g/\kappa\ll1$)~\cite{Weiss2016}. Therefore a full understanding of synchronization in OMSs in the presence of noise is needed.

For this purpose, in this paper, we explore the dynamics of a two-membrane OMS by generalizing the analysis of Holmes \textit{et al}.~\cite{Holmes2012} by including noise.  We apply stochastic Langevin equations to describe the system dynamics and simulate them numerically up to the long-time regime of $\sim 5$ mechanical relaxation times.
We reproduce the results of Ref.~\cite{Holmes2012} in the noiseless case, which can be described in terms of an amplitude-dependent Kuramoto-like model. When thermal noise is considered, we find that phase diffusion occurs, so that the two oscillators' phase becomes completely undetermined in the long-time regime, even though phase diffusion is significantly slowed down for increasing power of the drive. In fact, in the strong driving regime, the oscillator state can remain in a Gaussian state for a very long time. In the weak drive regime instead, we find that noise may induce a bistable behavior, in which two different limit cycles for each oscillator coexist and are both synchronized with a different relative phase. In such a regime the phase space probability distribution is bimodal, corresponding to the statistical mixture of two limit cycles.  More generally, we find that synchronization in this system is always robust with respect to thermal noise. We also noticed that even before the transition to synchronization, the two oscillators show a strong phase correlation (phase locking) with a residual slow drift in time, which we visualize by plotting the phase space probability distribution of a given resonator conditioned to a fixed value of the phase of the other one.

This paper is organized as follows: In Sec.~\ref{System dynamics}, we  present the dynamics of the system, including the stochastic Langevin equations we adopted, and the corresponding slowly-varying amplitude equations obtained after neglecting fast oscillating terms. In. Sec.~\ref{Bright and dark mode analysis}, we analyze such dynamics in terms of effective mechanical bright and dark mode in our system. In Sec.~\ref{The numerical analysis and synchronization measure}, we introduce the numerical methods and synchronization measures we used in this paper. In Sec.~\ref{Synchronization in the two-membranes optomechanical systems} we study in detail the synchronization phase diagram in the noiseless case, in Sec.~\ref{Multi-stability and phase diffusion induced by thermal noise} the noise induced non-Gaussian dynamics, i.e., phase diffusion and multistability, and in Sec.~\ref{Robustness with respect to thermal noise}, the robustness of synchronization with respect to thermal noise, and the presence of strong phase correlations between the two oscillators. Concluding remarks are given in the last section.

\section{System dynamics}  
\label{System dynamics}
\begin{figure}[]
\centering
\includegraphics[width=3.5in]{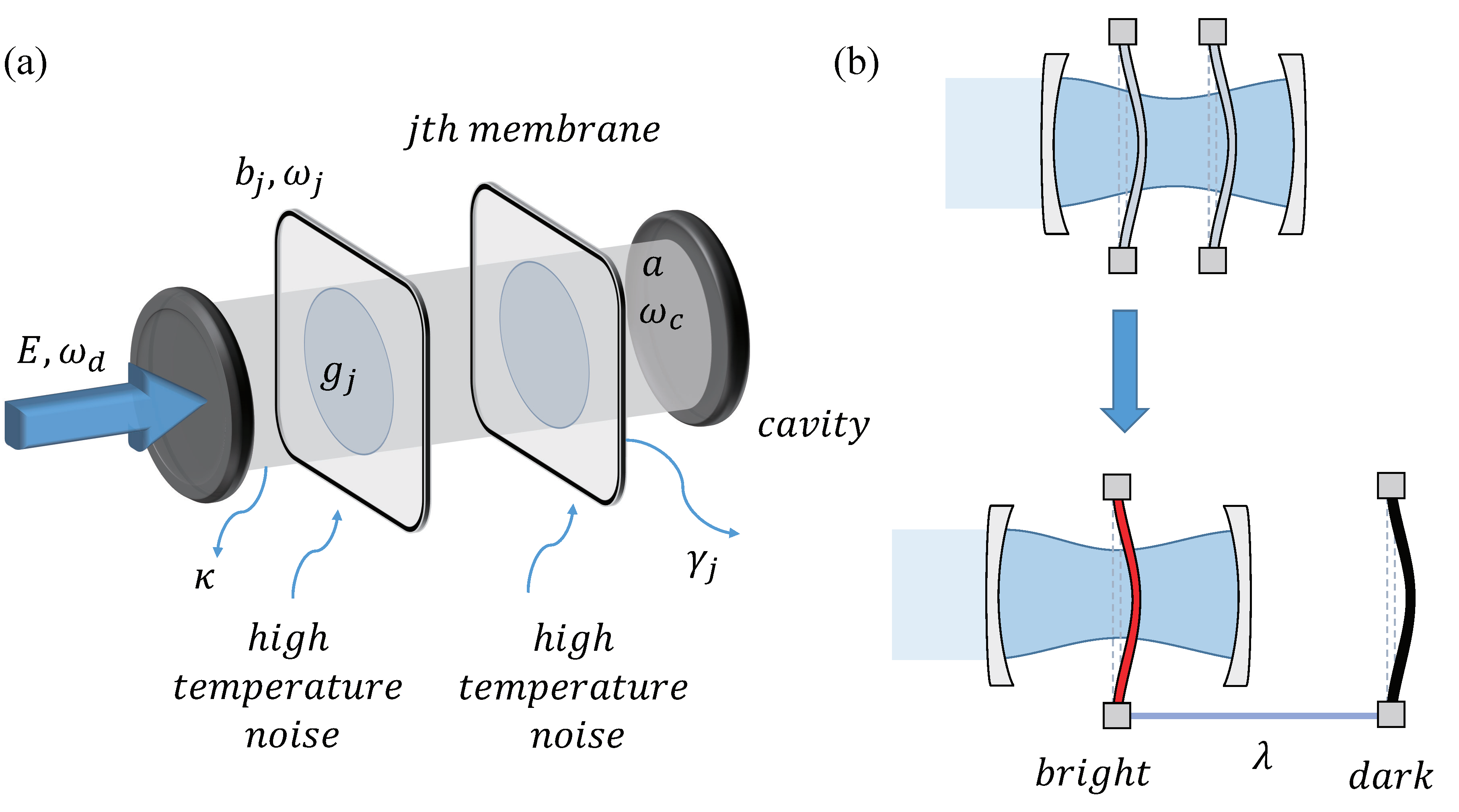}
\caption{ (a) Schematic diagram of a two-membrane OMS. (b) Our model can be also described by a ``bright'' mode coupled to the cavity field and a ``dark'' mode which  is decoupled from the cavity field. Here $\lambda$ denotes the effective coupling between the two modes.
\label{fig:1}}
\end{figure}
We consider two mechanical resonators coupled to a high finesse Fabry-P\'erot cavity driven by a pump laser beam, with input power $P$ and frequency $\omega_d$, different from the cavity mode frequency $\omega_c$ (see Fig.~1). In the frame rotating at the laser frequency, the system Hamiltonian reads ($\hbar=1$)
\begin{equation}
\begin{split}
H=&-\Delta a^\dagger a+iE(a^\dagger-a)\\
&+\sum_{j=1,2}\left[\omega_{j}b^\dagger_{j} b_{j}-g_{j}a^\dagger a(b^\dagger_{j}+b_{j})\right],
\end{split}
\label{eq:system Hamilton}
\end{equation}
where $a$ and $b_j$ are the optical and mechanical annihilation operators, $\Delta=\omega_d-\omega_c$, $\omega_j$ is the resonance frequency of the $j$-th mechanical resonator, with $g_j$ the corresponding single-photon optomechanical coupling rate, and $E=\sqrt{2\kappa_{in} P/\hbar\omega_d}$, with $\kappa_{in}$ the cavity field decay rate through the input port. The mechanical resonators and the cavity mode are coupled to their corresponding thermal reservoir at temperature $T$ through fluctuation-dissipation processes, which we include in the Heisenberg picture by adding dissipative and noise terms, yielding the following quantum Langevin equations~\cite{Giovannetti2001,Aspelmeyer2014}
\begin{subequations}
\begin{align}
\dot{a}=&(-\kappa+i\Delta)a+E\nonumber \\
&+\sum_{j=1,2} ig_j(b_j+b^\dagger_j)a+\sqrt{2\kappa_{in}}a^{in}+\sqrt{2\kappa_{ex}}a^{ex}, \label{eq:quantum langevin1}\\
\dot{b}_j=&(-\gamma_j-i\omega_j) b_j+ig_ja^\dagger a+\sqrt{2\gamma_j}b^{in}_j,
\label{eq:quantum langevin2}
\end{align}
\end{subequations}
where $\kappa=\kappa_{in}+\kappa_{ex}$ is the total cavity amplitude decay rate, $\kappa_{ex}$ is the optical loss rate through all the ports different from the input one, and $\gamma_j$ is the mechanical amplitude decay rate of oscillator $j$. $a^{in}(t)$, $a^{ex}(t)$  and $b^{in}_j$ are the corresponding noise reservoir operators, which are all uncorrelated from each other and can be assumed as usual to be Gaussian and white. In fact, they possess the correlation functions $\langle f(t)^\dagger f(t')\rangle = \bar{n}_f\delta(t-t')$ and $\langle f(t) f(t')^\dagger\rangle=(\bar{n}_f+1)\delta(t-t')$ where $f(t)$ is either $a^{in}(t)$, $a^{ex}(t)$ or $b^{in}_j$, and $\bar{n}_f=[\exp(\hbar\omega_f/k_bT)-1]^{-1}$ is the mean thermal excitation number for the corresponding mode.

In order to be more general and for a better comparison with previous works, we have assumed up to now a quantum description. However we shall restrict in this paper to study synchronization at room temperature $T \simeq 300$~K only, which justifies a classical treatment of the above Langevin equations
and implies a different treatment of optical and mechanical noise terms.
In fact, at optical frequencies $\omega_f/2\pi =\omega_c/2\pi \simeq 10^{14}$~Hz, so that $\bar{n}_f \simeq 0$, while at mechanical frequencies $\omega_f/2\pi =\omega_1/2\pi \simeq \omega_2/2\pi \simeq 10^{6}$ Hz implying $\bar{n}_f \simeq k_b T/\hbar\omega_1 \gg 1$. As a consequence, we expect that thermal noise will be dominant for the mechanical modes, but for large enough driving powers we cannot exclude in general the presence of non-negligible effects of the fluctuations of the intracavity field, due either to technical laser noise or ultimately to vacuum fluctuations.
Therefore we consider \emph{classical} complex random noises,  $\beta^{in}_j(t)$, $j=1,2$ (replacing the mechanical quantum thermal noise $b^{in}_j(t)$), and $\alpha^{opt}(t)$ (replacing the sum of optical vacuum noises $\sqrt{\kappa_{in}/\kappa} a^{in}(t)+\sqrt{\kappa_{ex}/\kappa}a^{ex}(t)$), with correlation functions
\begin{subequations}
\begin{align} \label{corre1}
&\langle \beta^{in}_j(t) \beta^{in}_{j'}(t')\rangle = \langle \alpha^{opt}(t) \alpha^{opt}(t')\rangle = 0, \\
&\langle \beta^{in,*}_j(t) \beta^{in}_{j'}(t')\rangle =(\bar{n}_b+1/2) \delta_{jj'}\delta(t-t'), \label{corre2}\\
&\langle \alpha^{opt,*}(t) \alpha^{opt}(t')\rangle=(1/2)\delta(t-t'), \label{corre3}
\end{align}
\end{subequations}
and we also have $\langle \beta^{in}_{j'}(t')\beta^{in,*}_j(t)\rangle=\langle \beta^{in,*}_j(t) \beta^{in}_{j'}(t')\rangle$ and $\langle \alpha^{opt,*}(t) \alpha^{opt}(t')\rangle=\langle \alpha^{opt}(t') \alpha^{opt,*}(t)\rangle$ because the $c$-numbers lose the commutation relation~\cite{Weiss2016,Li2017}. The quantum Langevin equations of Eqs.~(\ref{eq:quantum langevin1})-(\ref{eq:quantum langevin2}) are therefore well approximated by the set of coupled classical Langevin equations for the corresponding optical and mechanical complex amplitudes $\alpha(t)$ and $\beta_j(t)$~\cite{Weiss2016,Li2017,Wang2014},
\begin{subequations}
\begin{align}  \label{eq:c langevin1}
\dot{\alpha}(t)=&(i\Delta-\kappa)\alpha(t)+E\nonumber \\
&+\sum_{j=1,2} 2ig_j\text{Re}[\beta_j(t)]\alpha(t)+\sqrt{2\kappa}\alpha^{opt}(t),\\
\dot{\beta}_j(t)=&(-i\omega_j-\gamma_j)\beta_j(t)\nonumber\\
&+ig_j\vert\alpha(t)\vert^2+\sqrt{2\gamma_j}\beta^{in}_j(t).
\label{eq:c langevin}
\end{align}
\end{subequations}

In this paper we want to study the effect of noise on the synchronization of the two mechanical resonators realized by the optomechanical interaction with the same driven optical cavity mode, by generalizing the analysis of Ref.~\cite{Holmes2012}. With respect to Ref.~\cite{Holmes2012} we consider only the case of two different resonators within the cavity, which is however the experimentally relevant one (see Refs.~\cite{Piergentili2018,Gartner2018,Wei2019,Naesby2019}). In this system, under appropriate parameter regimes, the driven cavity mode sets each oscillator into a self-sustained limit cycle~\cite{Marquardt2006}, which may eventually become synchronized to each other. Synchronization may occur on a long timescale, determined by the inverse of the typically small parameters $\Delta \omega = \omega_2-\omega_1$ (never larger than 1 kHz), and $\gamma_j$ (order of Hz). Therefore it is physically useful to derive from the full dynamics of the classical Langevin equations (\ref{eq:c langevin1})-(\ref{eq:c langevin}), approximate equations able to correctly describe the slow, long time dynamics of the two mechanical resonators, leading eventually to synchronization.

We adapt here the slowly varying amplitude equations approach of Ref.~\cite{Holmes2012} to the case with noise studied here. Discarding here the limiting case of chaotic motion of the two resonators, which however occurs only at extremely large driving powers which are not physically meaningful for the Fabry-Perot cavity system considered here, it is known that each mechanical resonator, after an initial transient regime, sets itself into a dynamics of the following form~\cite{Marquardt2006}
\begin{equation}\label{ansatz}
\beta_j(t)=\beta_{0,j}+A_j(t)e^{-i\bar{\omega}t},
\end{equation}
where $\beta_{0,j}$ are constant, $A_j(t)$ are slowly-varying complex amplitudes of the oscillators, and $\bar{\omega}=(\omega_1+\omega_2)/2 \gg \Delta \omega$ is the average mechanical frequency. Eq.~(\ref{ansatz}) implies that we will study the long-time dynamics of the two mechanical resonators in the frame rotating at the fast reference frequency $\bar{\omega}$.
Inserting Eq.~(\ref{ansatz}) into Eq.~\eqref{eq:c langevin1}, and solving it formally by neglecting the transient term related to the initial value $\alpha(0)$, we have
\begin{equation}
\begin{split}
\alpha(t)=&\int_0^tdt'\left\{e^{\mathcal{L}(t-t')}[E+\sqrt{2\kappa}\alpha^{opt}(t')]\right. \\
& \left. \times \exp\left[2i g_q \int_{t'}^t dt'' \vert A_b(t'')\vert\cos(\bar{\omega}t''-\theta)\right] \right\},
\end{split}
\label{eq:cavity field1}
\end{equation}
where $\mathcal{L}=i [\Delta+\sum g_j(\beta^i_{0,j}+\beta^{i,*}_{0,j})]-\kappa$, $g_q=\sqrt{g_1^2+g_2^2}$, and we have defined the ``bright'' complex amplitude $A_b(t)=\vert A_b(t)\vert e^{i\theta}=g_q^{-1}\sum g_jA_j(t)$.

The amplitude $A_b(t)$ is much slower than the fast oscillations at $\bar{\omega}$ and one can treat it as a constant in the integral over $t''$ in Eq.~(\ref{eq:cavity field1}). Performing explicitly this integral one gets
\begin{equation}
\begin{split}
\alpha(t)&=e^{i\psi(t)}\int_0^tdt' e^{\mathcal{L}(t-t')}[E+\sqrt{2\kappa}\alpha^{opt}(t')]e^{-i\psi(t')}\\
& \equiv \alpha_E(t) + \delta \alpha (t),
\end{split}
\label{eq:cavity field2}
\end{equation}
where $\psi(t)=\xi\sin(\bar{\omega}t-\theta)$, with $\xi=2g_q\vert A_b\vert /\bar{\omega}$, and we have defined the intracavity field $\alpha_E(t)$ proportional to driving rate $E$ and $\delta \alpha (t)$ related to the input noise $\alpha^{opt}(t)$.

For the intracavity amplitude $\alpha_E(t)$ we follow the usual approach~\cite{Marquardt2006,Holmes2012} and use the Jacobi-Anger expansion for the $e^{-i\psi(t')}$ factor within the integral, i.e., $e^{-i\xi\sin\phi}=\sum_n J_n(-\xi)e^{in\phi}$, ($\phi = \bar{\omega}t'-\theta$ and $J_n$ is the $n$-th Bessel function of the first kind), and finally get for the intracavity field amplitude
\begin{equation}
 \alpha_E(t)  =E e^{i\psi(t)}\sum_{n=-\infty}^{\infty}\dfrac{J_n\left(-\xi\right)e^{in(\bar{\omega}t-\theta)}}{in\bar{\omega}-\mathcal{L}}.
\label{eq:cavity field solution}
\end{equation}
For the fluctuation term we notice instead that, due to Eqs.~(\ref{corre1}), (\ref{corre3}), $ \alpha^{opt}(t)e^{-i\psi(t)}$ possesses the same correlation functions of $ \alpha^{opt}(t)$ and therefore the factor $e^{-i\psi(t')}$ can be practically neglected in the integral, and we have simply
\begin{equation}
\delta \alpha(t) =\sqrt{2\kappa} e^{i\psi(t)}\int_0^tdt' e^{\mathcal{L}(t-t')}\alpha^{opt}(t').
\label{eq:cavity field fluctuation}
\end{equation}
We have now to insert these expressions into the radiation pressure force term within Eq.~(\ref{eq:c langevin}) for the mechanical motion, and derive an equation for the unknown quantities $\beta^i_{0,j}$ and $A_j(t)$. Since the intracavity optical fluctuations are small, we can reasonably approximate the radiation pressure term at first order in $\delta \alpha(t)$,
\begin{equation}
ig_j \vert \alpha(t)\vert^2 \simeq ig_j\vert \alpha_E(t) \vert^2 + ig_j \eta_{opt}(t)
\label{eq:radpress}
\end{equation}
where
\begin{equation}\label{doublesum}
\vert \alpha_E(t) \vert^2 = E^2 \sum_{n,m=-\infty}^{\infty}\dfrac{J_n\left(-\xi\right)J_m\left(-\xi\right)e^{i(n-m)(\bar{\omega}t-\theta)}}{(in\bar{\omega}-\mathcal{L})(-im\bar{\omega}-\mathcal{L}^*)},
\end{equation}
and
\begin{equation}
\eta_{opt}(t) =\alpha_E(t) \delta \alpha^*(t)+ \alpha_E^*(t)\delta \alpha(t).
\end{equation}
Using the fact that $\beta_{0,j}$ are assumed constant, and neglecting all terms oscillating faster than $\bar{\omega}$, i.e., keeping only the resonant terms in Eq.~(\ref{doublesum}) [$n-m=0$ for $\beta_{0,j}$ and $n-m=-1$ for the amplitudes $A_j(t)$], we get
\begin{equation}
(\gamma_j+i\omega_j) \beta_{0,j}=ig_j\sum_{n=-\infty}^{\infty}\dfrac{E^2J_n\left(-\xi\right)^2}{(in\bar{\omega}-\mathcal{L})(-in\bar{\omega}-\mathcal{L}^*)},
\label{eq:zero-order steady}
\end{equation}
for $\beta_{0,j}$, and
\begin{equation}
\begin{split}
\dot{A}_j(t)=&\left[-\gamma_j-i(-1)^j\frac{\Delta\omega}{2}\right] A_j(t)\\ &+ig_je^{i\theta}E^2\sum_n\dfrac{J_n\left(-\xi\right)J_{n+1}\left(-\xi\right)}{[in\bar{\omega}-\mathcal{L}][-i(n+1)\bar{\omega}
-\mathcal{L}^*]}\\&+\sqrt{2\gamma_j}\beta^{in}_j(t)+ig_j\eta_{opt}(t),
\label{eq:first-order solution}
\end{split}
\end{equation}
for the slowly varying amplitudes $A_j(t)$. Eq.~(\ref{eq:zero-order steady}) cannot be easily used to determine the values of $\beta_{0,j}$ because its right hand side depends upon the slowly varying unknown variable $\xi$. Instead, we obtained the values of $\beta_{0,j}$ by solving numerically Eqs.~(\ref{eq:c langevin1})-(\ref{eq:c langevin}) without noise terms, and we used them to define the effective cavity detuning
\begin{equation}
\Delta \to \Delta_{eff}=\Delta +\sum g_j(\beta_{0,j}+\beta^{*}_{0,j}),
\end{equation}
which is the actual parameter controlled in an experiment. As a consequence, $\mathcal{L}=i\Delta_{eff}-\kappa$ becomes a given known parameter, and we have verified that Eq.~(\ref{eq:zero-order steady}) is self-consistently satisfied in the long-time limit when $\xi(t)$ reaches its stationary value.

Eq.~(\ref{eq:first-order solution}) can be rewritten in a better form by defining the following regular dimensionless auxiliary function $\mathcal{F}(|A_b|,\bar{\omega},\kappa,\Delta_{eff})$ as
\begin{equation}
\mathcal{F}=\dfrac{E^2}{\vert A_b\vert}\sum_{n=-\infty}^{\infty}\dfrac{J_n\left(-\xi\right)J_{n+1}\left(-\xi\right)}{[in\bar{\omega}-\mathcal{L}][-i(n+1)\bar{\omega}-\mathcal{L}^*]},
\label{eq:auxiliary function}
\end{equation}
the amplitude equations can be finally given as:
\begin{equation}
\begin{split}
\dot{A}_j(t)=&\left[-\gamma_j-i(-1)^j\frac{\Delta\omega}{2}\right] A_j(t)+\sqrt{2\gamma_j}\beta^{in}_j(t)\\
&+ig_jA_b(t) \mathcal{F}(|A_b|,\bar{\omega},\kappa,\Delta_{eff})+ig_j \eta_{opt}(t).
\label{eq:first-order finnal}
\end{split}
\end{equation}

\section{Bright and dark mode analysis}
\label{Bright and dark mode analysis}
Eq.~(\ref{eq:first-order finnal}) does not have only the advantage of providing a useful tool for the long-time numerical simulation of the problem, but also suggests a simpler approach for better understanding the physics of the system when looking for synchronization of the two mechanical resonators. In fact, as we have seen, the amplitude variable $A_b(t)\propto g_1A_1+g_2 A_2$, which we call bright because it is the one directly interacting with the cavity mode plays an important role in the equations. It is convenient to directly write the evolution equation in terms of $A_b(t)$ and of an independent, orthogonal variable, which we call ``dark'' mode
\begin{equation}\label{darkmode}
  A_d(t)=\frac{g_1 A_2(t)-g_2 A_1(t)}{g_q},
\end{equation}
so that the relation between the original amplitudes associated with each mechanical resonator and the bright and dark ones are, in fact, a coordinate rotation by an angle $\theta_{rot}$ such that $\tan \theta_{rot}=g_2/g_1$. As a consequence, the inverse relations are
\begin{subequations}
\begin{align}
A_1(t)=& \frac{g_1 A_b(t)-g_2 A_d(t)}{g_q}, \label{inverse1}\\
A_2(t)=& \frac{g_1 A_d(t)+g_2 A_b(t)}{g_q}.\label{inverse2}
\end{align}
\end{subequations}
After some lengthy but straightforward algebra, we get the following equations for the new amplitude variables
\begin{subequations}
\begin{align}
\dot{A}_b(t)=&-\Gamma_b A_b(t)+ig_q\mathcal{F}(|A_b|)A_b(t)\nonumber\\&-\lambda A_d(t) +\beta^{in}_b(t)+i g_q \eta_{opt}(t), \label{eq:bright}\\
\dot{A}_d(t)=&-\Gamma_d A_d(t)-\lambda A_b(t) +\beta^{in}_d(t), \label{eq:dark}
\end{align}
\end{subequations}
where we have omitted for simplicity the dependence of $\mathcal{F}$ upon the other parameters. Using the definition $\Delta \gamma =\gamma_2-\gamma_1$, the coefficients appearing in these coupled equations are the coupling between dark and bright mode
\begin{equation}\label{dark-bright-coup}
  \lambda=\frac{g_1g_2}{g_q^2}(\Delta \gamma+i \Delta \omega),
\end{equation}
and the two complex rates
\begin{equation}\label{rates}
  \Gamma_{b/d}=\frac{\gamma_1+\gamma_2}{2}\pm \frac{\lambda}{2}\frac{g_2^2-g_1^2}{g_1g_2},
\end{equation}
and we have defined the corresponding new thermal noise terms
\begin{subequations}
\begin{align}
  \beta^{in}_b(t)&= \frac{\sqrt{2\gamma_1} g_1 \beta^{in}_1(t)+\sqrt{2\gamma_2} g_2 \beta^{in}_2(t)}{g_q}, \label{noiseb}\\
\beta^{in}_d(t)&= \frac{\sqrt{2\gamma_2} g_1 \beta^{in}_2(t)-\sqrt{2\gamma_1} g_2 \beta^{in}_1(t)}{g_q}.\label{noised}
\end{align}
\end{subequations}
These noise terms have the following correlation functions
\begin{subequations}
\begin{align}
 \langle \beta^{in,*}_b(t) \beta^{in}_{b}(t')\rangle &= \frac{2\gamma_1 g_1^2 \bar{n}_1+2\gamma_2 g_2^2 \bar{n}_2}{g_q^2}\delta(t-t'),  \label{correb}\\
\langle \beta^{in,*}_d(t) \beta^{in}_{d}(t')\rangle &= \frac{2\gamma_1 g_2^2 \bar{n}_1+2\gamma_2 g_1^2 \bar{n}_2}{g_q^2}\delta(t-t').\label{corred}
\end{align}
\end{subequations}
Notice that these two effective thermal noise terms are correlated in general, since it is
\begin{equation}
  \langle \beta^{in,*}_b(t) \beta^{in}_{d}(t')\rangle =  \frac{2g_1g_2}{g_q^2}(\gamma_1 \bar{n}_1-\gamma_2 \bar{n}_2)\delta(t-t'). \label{correcorre}
\end{equation}
The definitions of bright and dark modes are evident from Eqs.~\eqref{eq:bright}-\eqref{eq:dark}: only $A_b(t)$ is directly coupled to the cavity mode via the nonlinear term $ig_q \mathcal{F}(|A_b|)A_b(t)$, while $A_d(t)$ feels the effect of radiation pressure only via its coupling with the bright mode, which is zero in the case of identical mechanical resonators. Moreover the optical noise $\eta_{opt}$ affects only the bright mode.

The simple form of the dynamical equations for the bright and dark mode suggests a general way for the formal solution of the problem. Since we are interested in the very long time dynamics, we neglect the transient term associated with $A_d(0)$ and we first write the formal solution for $A_d(t)$ as a function of $A_b(t)$,
\begin{equation}
  A_d(t) =  \int_0^t dt' e^{-\Gamma_d(t-t')}[-\lambda A_b(t')+\beta^{in}_d(t')] , \label{formal1}
\end{equation}
and then replace it within the equation for $A_b(t)$, yielding the following integro-differential equation for the dynamics of the bright mode amplitude alone,
\begin{equation}
\begin{split}
\dot{A}_b(t)=&-\Gamma_b A_b(t)+\lambda^2 \int_0^t dt' e^{-\Gamma_d(t-t')}A_b(t') \\
&+ig_q\mathcal{F}(|A_b|)A_b(t)+\beta^{in}_b(t)+i g_q \eta_{opt}(t)\\&-\lambda  \int_0^t dt' e^{-\Gamma_d(t-t')}\beta^{in}_d(t'),
\end{split}
\end{equation}
Formally the problem could be exactly solved by first solving this latter integro-differential equation for $A_b(t)$, then using this solution within Eq.~\eqref{formal1} in order to get $A_d(t)$ and finally get the exact form for $A_1(t)$ and $A_2(t)$ using the change of variables of Eqs.~\eqref{inverse1}-\eqref{inverse2}.

\section{The numerical analysis and synchronization measure}
\label{The numerical analysis and synchronization measure}
In this section, we describe the numerical analysis employed here, and the physical quantities adopted to quantify synchronization and more in general the dynamical behavior of the two mechanical resonators.
As noted above, Eq.~\eqref{eq:first-order finnal} provides with very good approximation the long-time dynamics of the two mechanical resonators, on times of the order of $\Delta \omega^{-1}$ and $\gamma_j^{-1}$, while the classical Langevin equations \eqref{eq:c langevin1}-\eqref{eq:c langevin} provide the full dynamical evolution also at the much faster timescales $\kappa^{-1}$ and $\bar{\omega}^{-1}$. This full dynamics is however relevant for the determination of the initial transient evolution of the two coupled mechanical resonators. Therefore we need to provide the correct initial conditions for the slowly-varying amplitude equations of Eq.~\eqref{eq:first-order finnal}. Our simulation process followed these steps:
\begin{itemize}
  \item [\romannumeral1)]
 Numerically solve the full classical equations Eqs.~\eqref{eq:c langevin1}-\eqref{eq:c langevin} without noise up to a long time $t_c$ of the order of few $\gamma_j$ to get the frequency modification term $\beta_{0,j}$, and substitute the result in Eq.~\eqref{eq:zero-order steady} to verify its accuracy.
  \item [\romannumeral2)]
Simulate the full classical Langevin equations \eqref{eq:c langevin1}-\eqref{eq:c langevin} for a time interval $t_1$ longer than the fast timescale $\simeq 1/\kappa$ and shorter than the slow timescale $\simeq 1/\gamma_j$, and record the final states of two oscillators. Those final states will be taken as the initial states for the next stage.
  \item [\romannumeral3)]
Simulate Eq.~\eqref{eq:first-order finnal} for a time interval $t_2$ starting from the initial conditions obtained in Step \romannumeral2). The frequency modification in Eq.~\eqref{eq:first-order finnal} is obtained by Step \romannumeral1). The auxiliary function $\mathcal{F}$ of Eq.~(\ref{eq:auxiliary function}) has been evaluated summing over the index $n$ in the interval $n\in[-100,100]$.
\end{itemize}
We typically average over $N$ trajectories,
each starting from a random initial condition for $\alpha(0)$ and $\beta_j(0)$, chosen from the zero-mean Gaussian distribution associated with the corresponding initial thermal state, i.e., the vacuum state for the optical mode, and the thermal state with $\bar{n}_j \simeq k_b T/\hbar\omega_j \gg 1$ for the mechanical resonators. Of course, each trajectory employs a different realization of the Gaussian noises involved, $\beta_j^{in}$ and $\alpha^{opt}$.

Moreover, we focus our numerical study onto a realistic scenario at room temperature, which is the most interesting one for applications, and consider the set of parameters of Ref.~\cite{Piergentili2018}, that is $\omega_1/2\pi=235.810$\,kHz, $\omega_2/2\pi=236.580$\,kHz, $g_1/2\pi=0.3$\,Hz, $g_2/2\pi=0.28$\,Hz, $\gamma_1/2\pi=1.64$\,Hz and $\gamma_2/2\pi=9.37$\,Hz. We then take as variable parameters the detuning $\Delta$, cavity decay rate $\kappa$ and the driving rate $E$, which is equivalent to change the input power $P$. The tiny difference in phonon number caused by $\omega_1\neq\omega_2$ is neglected, so that we set $\bar{n}_1=\bar{n}_2=2.5\times10^7$ corresponding to the room temperature case ($T\sim 300$\,K).

With the chosen set of parameters, we can safely neglect the effect of optical vacuum noise on the synchronization dynamics of the mechanical resonators, i.e., we can neglect $\eta_{opt}$ within Eq.~\eqref{eq:first-order finnal} and therefore also $\alpha^{opt}$ within Eq.~\eqref{eq:c langevin1}. In fact, one can easily see that while the effects of $\beta_j^{in}$ scale with $\gamma_j \bar{n}_j$, those of $\eta_{opt}$ scale with $g_j^2 |\langle \alpha \rangle |^2/\kappa$. Therefore, the effects of thermal and optical vacuum noises are comparable only when the cooperativity $C_j = g_j^2 |\langle \alpha \rangle |^2/\kappa \gamma_j $ is comparable to the mean thermal phonon number $\bar{n}_j$. At room temperature and weak optomechanical coupling conditions chosen above, this condition is always far from being satisfied, even when considering quite unrealistic very large input powers of hundreds of mW. Therefore we will drop the noise term $\eta_{opt}$ from now on.

Phase synchronization is generally measured by means of the Pearson's correlation coefficient, expressed in the more general case where chaotic motion can be present, as~\cite{Giorgi2012,Giorgi2013,Li2017,Manzano2013}:
\begin{equation}
\begin{split}
\mathcal{C}[f,g](t,\Delta t)=\dfrac{\overline{\delta f\delta g}}{\sqrt{\overline{\delta f^2}\times\overline{\delta g^2}}},
\label{eq:Pearson}
\end{split}
\end{equation}
where $\overline{o}=\Delta t^{-1}\int^{t+\Delta t}_{t}o(\tau)d\tau$ and $\delta o=\overline{o}-o$; we choose for $f$ and $g$ the dynamical quantities Re$(A_1)=|A_1|\cos\theta_1$ and Re$(A_2)=|A_2|\cos\theta_2$. When the system does not exhibit chaotic behavior, we can also characterize synchronization in terms of the phase difference~\cite{Weiss2016},
\begin{equation}
\begin{split}
\mathcal{P}(t)=\cos\theta_-=\cos(\theta_1-\theta_2).
\label{eq:phase error measure}
\end{split}
\end{equation}
In the absence of noise, these two quantities evaluated after a transient regime provide a direct measure of synchronization. In the presence of noise instead, consistent stable results are obtained only after appropriate averaging.
More precisely, for the $i$-th stochastic trajectory generated in the simulation we record the result as $\mathcal{M}^i(t)$, where $\mathcal{M}$ is either $\mathcal{C}$ or $\mathcal{P}$, and we first perform an ensemble average of these synchronization measures~\cite{Li2017},
\begin{equation}
\begin{split}
\mathcal{M}(t)=\dfrac{1}{N}\sum_{i=1}^{N} \mathcal{M}^i(t).
\label{eq:m measure}
\end{split}
\end{equation}
Then we perform a time average of the above quantity, that is,
\begin{equation}
\begin{split}
\bar{\mathcal{M}}=\dfrac{1}{T}\int\mathcal{M}(t)dt,
\label{eq:m time measure}
\end{split}
\end{equation}
where $T$ is a large enough time interval ensuring stable values. The two measures provide a very similar description of synchronization, and we notice that both measures yield $\bar{\mathcal{M}}=1$, $0$ and $-1$ when the system is $0$-phase synchronized, un-synchronized, and $\pi$-phase synchronized, respectively.

We will also numerically study phase diffusion for each resonator, which we will quantify in terms of the following standard deviation averaged over the $N$ trajectories
\begin{equation}
\begin{split}
\mathcal{S}_j= \sqrt{\frac{2}{N}\sum_i \left[\cos(\arg[A'^i_j])-\frac{1}{N}\sum_i\cos( \arg[A'^i_j])\right]^2}
\end{split}
\label{eq: Phase fluctuation}
\end{equation}
where $A'^i_j=A^i_je^{-i\arg[\langle A^i_j \rangle ]}$ is the amplitude of the j-th oscillator in the i-th trajectory defined with respect to a reference frame rotating with the phase of the average trajectories, as suggested in Ref.~\cite{Mari2013}. The cosine function is introduced to eliminate the multiple values of the phase, and the normalization factor $2$ here ensures that a completely homogeneous phase distribution over $2 \pi$ corresponds to $\mathcal{S}=1$.

Finally we will also characterize in a more visual way synchronization and phase correlations in terms of probability distributions. In particular we will consider the probability distribution of the phase difference $\theta_-$~\cite{Lee2013,Jessop2019,Lorch2017}, evaluated numerically as
\begin{equation}
\begin{split}
P_{\theta_-}(\theta)=\lim_{h\rightarrow 0}\dfrac{N_{\theta}}{Nh},
\end{split}
\label{eq:phase distribution function}
\end{equation}
where $N_{\theta}$ is the number of $\theta^i_-$ satisfying $\theta^i_-\in(\theta-h/2,\theta+h/2]$. Then we will also plot the reduced Wigner function of each mechanical oscillator, which in the classical regime considered here does not assume negative values~\cite{book}, and is just a standard phase-space probability distribution, which can be evaluated as
\begin{equation}
\begin{split}
W_j(Q ,P)=\lim_{h\rightarrow 0}\dfrac{N_{Q,P}}{N h^2},
\end{split}
\label{eq:phase distribution function}
\end{equation}
where $N_{Q,P}$ is the number of results satisfying $Q_j^i\in(Q-h/2,Q+h/2]$ and $P_j^i\in(P-h/2,P+h/2]$, with $Q_j^i=A^i_j+A^{i*}_j$ and $P_j^i=i(A_j^{i*}-A_j^i)$ the two resonator quadratures.

\section{Synchronization phase diagram in the noiseless case}
\label{Synchronization in the two-membranes optomechanical systems}
Our study will first review the mean-field case in which noise is neglected, and provide the synchronization phase diagram as a function of the relevant parameters, i.e., driving strength $E$, cavity detuning $\Delta$, and cavity decay $\kappa$. In the subsequent subsections we will discuss the influence of noise in the different parameter regimes.

\begin{figure}[]
\centering
\includegraphics[width=3.5in]{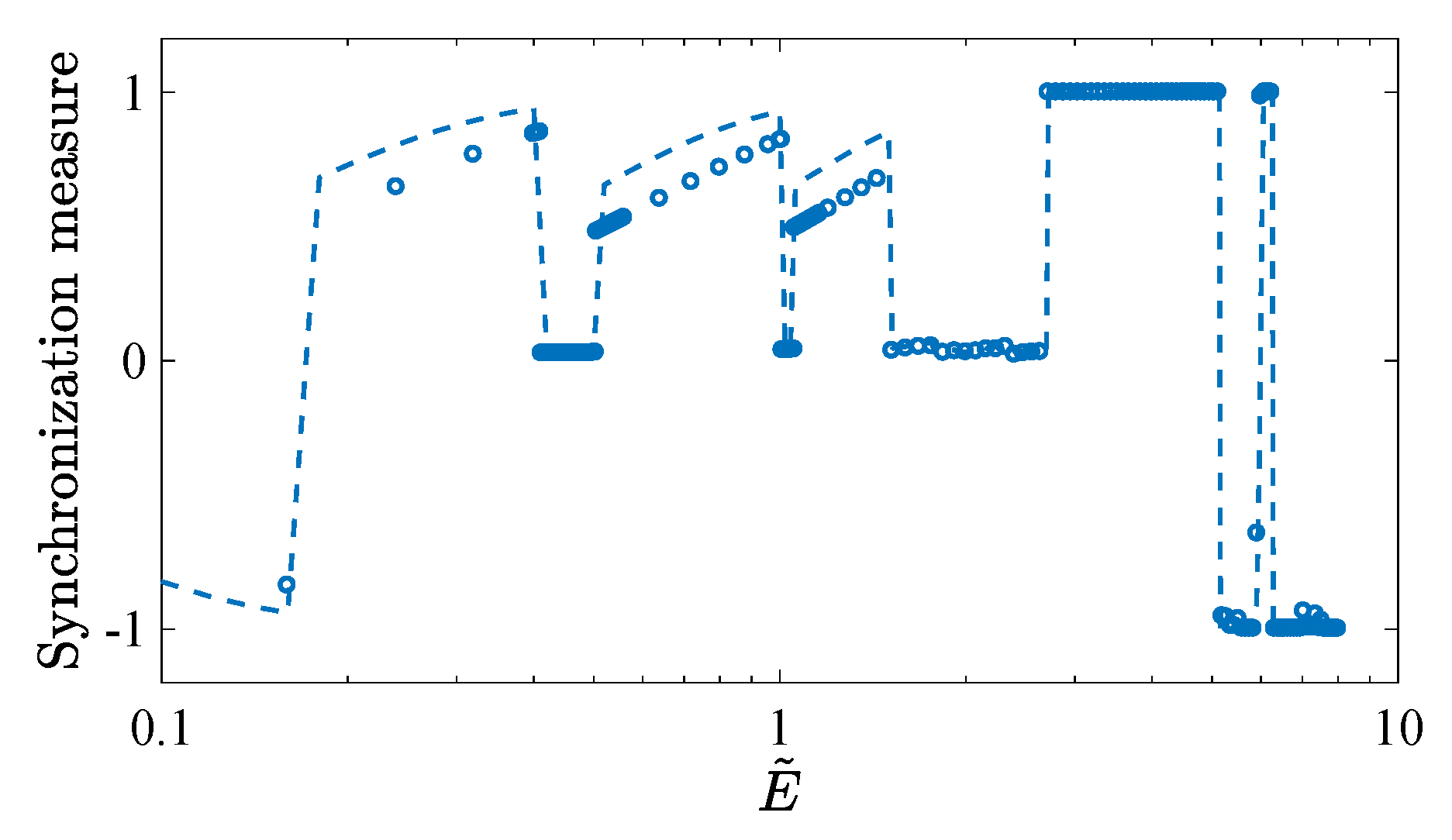}
\caption{Synchronization measure as a function of the dimensionless driving amplitude $\tilde{E}=E/E_s$, where $E_s = 10^5 \omega_1$ corresponds to an input power $P \simeq 5.5$ mW in the case of the experimental parameter regime of Ref.~\protect\cite{Piergentili2018}. The blue points are obtained with the full classical Langevin equations of Eqs.~(\protect\ref{eq:c langevin1})-(\protect\ref{eq:c langevin}), and evaluating the Pearson's correlation coefficient $\mathcal{C}$, averaged over the time interval $t\in[0.34\text{s},0.41\text{s}]$. The blue dashed line is instead calculated by simulating the amplitude equation~(\protect\ref{eq:first-order finnal}) and choosing $\mathcal{P}$, averaged over the time interval $t\in[0.20\text{s},0.41\text{s}]$ as synchronization measure.
We have chosen the following parameters $\Delta=\Delta_s=\bar{\omega}$, $\kappa=\kappa_s=\bar{\omega}/2$, while the other parameters have been given in the main text and correspond to those of Ref.~\protect\cite{Piergentili2018}.
\label{fig:2}}
\end{figure}
As a first preliminary step we have verified for a wide range of parameters that the long-time predictions of the slowly-varying amplitude equation~(\ref{eq:first-order finnal}) and of the full classical Langevin equations~(\ref{eq:c langevin1})-(\ref{eq:c langevin}) coincide in the noiseless case. The expected agreement between the two approaches is shown in Fig.~\ref{fig:2}, where we plot the behavior of the long-time synchronization measure as a function of the dimensionless driving amplitude $\tilde{E}=E/E_s$, where $E_s = 10^5 \omega_1$ corresponds to an input power $P \simeq 5.5$ mW in the case of the experimental parameter regime of Ref.~\cite{Piergentili2018}. The blue points have been obtained with the full classical Langevin equations of Eqs.~(\ref{eq:c langevin1})-(\ref{eq:c langevin}), and quantifying synchronization with the Pearson's correlation coefficient $\mathcal{C}$, averaged over the time interval $t\in[0.34\text{s},0.41\text{s}]$. The blue dashed line is instead calculated by simulating the amplitude equations (\ref{eq:first-order finnal}) and choosing $\mathcal{P}=\cos \theta_-$, averaged over the time interval $t\in[0.20\text{s},0.41\text{s}]$ as synchronization measure. In this latter case we started the time average at an earlier time to double-check the correctness of the choice of the initial conditions for the amplitude equations in Step iii) of Sec. IV.
We find that two methods are in good agreement with each other over a wide interval of $E$, and that this remains true regardless the adopted synchronization measure, $\mathcal{C}$ or $\mathcal{P}$.
As described in Sec. IV, when employing Eq.~(\ref{eq:first-order finnal}), we first solved the full Langevin
equations~(\ref{eq:c langevin1})-(\ref{eq:c langevin}) up to a time $t_1$ ($t_1=3\times10^5/\omega_1\sim 0.2$s) in order to get the correct initial state for the amplitude equations. We have seen that the system is not so sensitive to the initial state if the drive is not particularly strong ($\tilde{E}<4$), and that the initial conditions are correctly chosen even when $t_1$ is decreased down to $100/\omega_1$. Since $\tilde{E}>4$ corresponds to large and quite unrealistic input powers ($P > 88$ mW), we will not numerically study this parameter regime further.

\begin{figure}[]
\centering
\includegraphics[width=3.5in]{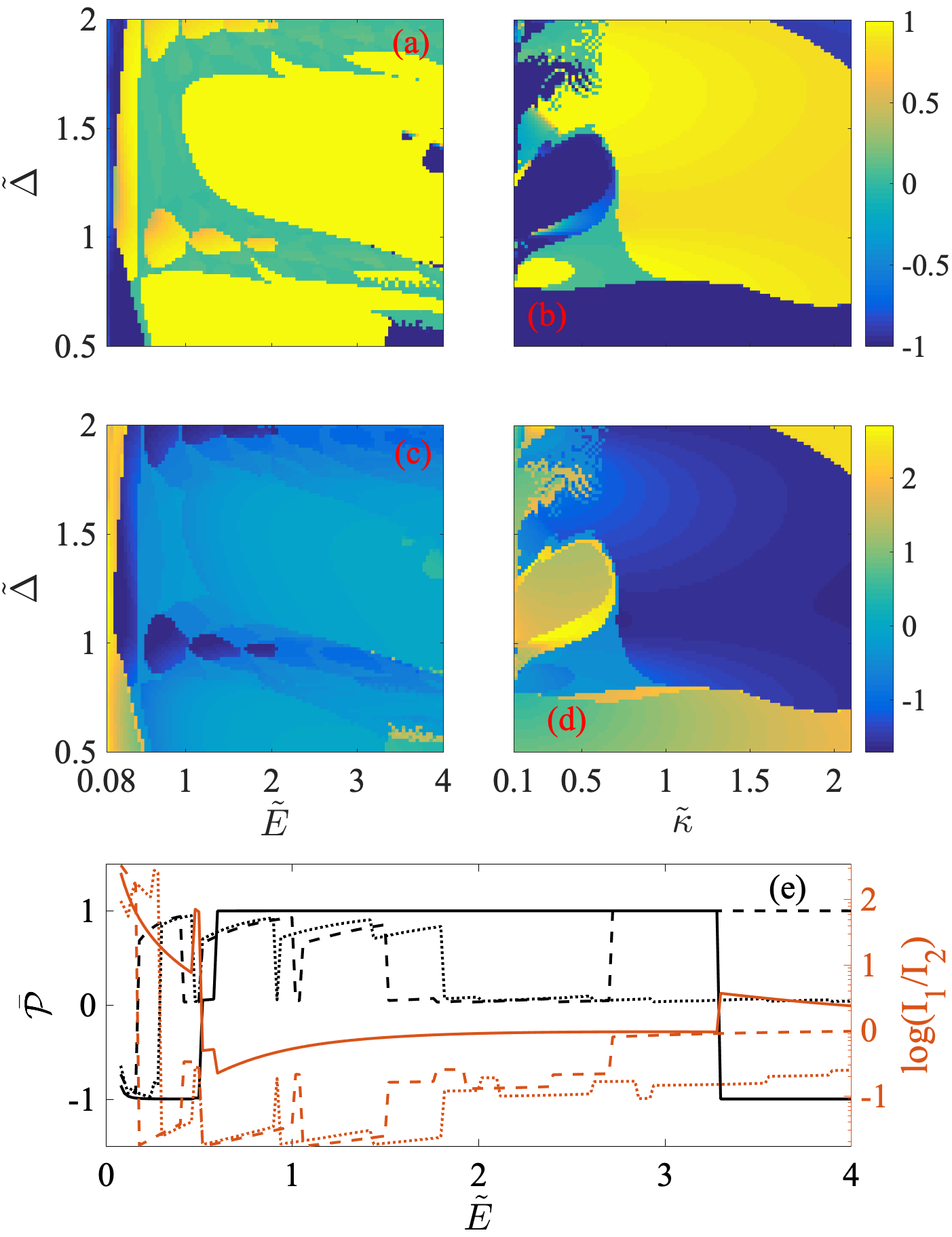}
\caption{(a) and (b): Synchronization phase diagrams in terms of $\mathcal{P}$ in the driving-detuning plane and detuning-cavity decay plane. Here the dimensionless variables are defined as $\tilde{\Delta}=\tilde{\Delta}/\Delta_s$, $\tilde{\kappa}=\kappa/\kappa_s$ respectively, with $\Delta_s=\bar{\omega}$, $\kappa_s=\bar{\omega}/2$. (c) and (d): Amplitude ratio [$\log_{10}(I_1/I_2)$] in the driving-detuning plane and detuning-cavity decay plane.  (e): Sectional view of (a) (black lines), and (c) (red lines), by fixing different values of the detuning $\tilde{\Delta}$. The solid lines, dashed lines and dotted lines correspond to the case $\tilde{\Delta}=0.5$, $1$ and $2$, respectively. The other parameters are the same as in Fig.~\ref{fig:2}.
\label{fig:3}}
\end{figure}
Fig.~\ref{fig:2} shows a series of synchronization crossover points, which encourages us to explore the synchronization scenario in more detail, extending in various directions the analysis of Ref.~\cite{Holmes2012}. In Fig.~\ref{fig:3}, we show the synchronization phase diagram (using $\mathcal{P}$) in the driving-detuning plane (a) and the detuning-cavity decay plane (b), respectively, obtained from the numerical solution of Eq.~(\ref{eq:first-order finnal}). It is evident that the present OMS offers a synchronization phase diagram much richer than the standard Kuramoto model. In order to explain the complex dynamics of the system, we rewrite the noiseless amplitude equation of Eq.~(\ref{eq:first-order finnal}) in terms of the modulus and phase of the two complex amplitudes, $A_j = I_j e^{i\theta_j}$,
 \begin{eqnarray}
&&\dot{I}_1=-\gamma_1 I_1-\frac{g^2_1}{g_q}I_1\mathcal{F}_i-\frac{g_1g_2}{g_q}I_2(\mathcal{F}_i\cos\theta_--\mathcal{F}_r\sin\theta_-),\nonumber \\
&&\dot{I}_2=-\gamma_2 I_2-\frac{g^2_2}{g_q}I_2\mathcal{F}_i-\frac{g_1g_2}{g_q}I_1(\mathcal{F}_i\cos\theta_-+\mathcal{F}_r\sin\theta_-),\nonumber \\
&&\dot{\theta}_-=-\Delta \omega-\frac{g^2_1-g^2_2}{g_q}\mathcal{F}_r \label{eq:c ap langevin}\\
&&-\frac{g_1g_2}{g_q}\left[\left(\dfrac{I_2}{I_1}-\dfrac{I_1}{I_2}\right)\mathcal{F}_r\cos\theta_-+\left(\dfrac{I_2}{I_1}+\dfrac{I_1}{I_2}\right)\mathcal{F}_i\sin\theta_-\right],
\nonumber
\end{eqnarray}
where $\mathcal{F}=\mathcal{F}_r+i\mathcal{F}_i$. Eq.~\eqref{eq:c ap langevin} demonstrates that the phase difference of the two oscillators obeys a Kuramoto-like equation. The main difference with the standard Kuramoto model is that Eq.~\eqref{eq:c ap langevin} includes a nonstandard $\cos \theta_-$ term, and that the coupling coefficients are not fixed but depend upon the moduli $I_j$, also through $\mathcal{F}_r$ and $\mathcal{F}_i$. The limit cycle dynamics ensure that $I_j$ assume stable values in the long-time regime, and therefore we can make a qualitative analysis of the synchronization phase diagram by regarding $I_1$ and $I_2$ as two given parameters. When the two limit cycles have comparable amplitudes, $I_1\simeq I_2$, the cosine term disappears, one has the standard Kuramoto model. The system will therefore achieve perfect $0$-phase synchronization in this case when $\vert g_q \Delta \omega-(g^2_1-g^2_2)\mathcal{F}_r\vert\leq 2g_1g_2|\mathcal{F}_i|$. When instead the two amplitudes are very different, the cosine term can shift the equilibrium position of $\theta_-$, thus causing the system to deviate from perfect phase synchronization, and even achieve $\pi$-synchronization.

We verify the above analysis and the presence of a strong similarity between the synchronization phase diagram and the behavior of the amplitude ratio $I_1/I_2$ by plotting the $\rm{log_{10}}$ of the latter in Fig.~\ref{fig:3}(c) and (d) for the same parameter regime of Fig.~\ref{fig:3}(a) and (b). The two contour plots show a remarkable similarity, and the transition from one synchronization phase to the other is always associated to a distinct jump in the value of $\log_{10}(I_1/I_2)$. This is more evident in Fig.~\ref{fig:3}(e), where $\tilde{\mathcal{P}}$ and $\log_{10}(I_1/I_2)$ are plotted as a function of the driving amplitude $\tilde{E}$ at three different values of $\tilde{\Delta}$. One can see that the behavior of the synchronization measure is very similar to that of the amplitude ratio. Specifically, for not too large drivings, the occurrence of the $0$/$\pi$-synchronization crossover is always accompanied by the transition from $I_1\gg I_2$ to $I_2\gg I_1$, which corresponds to the change of sign of the cosine term coefficient. Therefore, one can predict the synchronization behavior in this model by looking at the amplitude ratio of the two oscillators.

Finally we notice from Fig.~\ref{fig:3} that many synchronization crossovers occur at the first ($\tilde{\Delta}=1$) and at the second ($\tilde{\Delta}=2$) blue motional sidebands, associated with the presence of some small ``islands'' around these sidebands in Fig.~\ref{fig:3}(a). Physically, this phenomenon indicates that the driving field will enhance the nonlinear effects when it resonates with the sidebands. From Fig.~\ref{fig:3}(b) instead we see a complex synchronization phase diagram in the good cavity limit of smaller $\kappa$, which is associated with the fact that the radiation pressure nonlinearity has stronger effects when there are more photons in the cavity.

\section{Multi-stability and phase diffusion induced by thermal noise}
\label{Multi-stability and phase diffusion induced by thermal noise}
When the effects of thermal noise are taken into account, two non-Gaussian features that cannot be described by mean-field and simple linearization treatments are observed: i) phase diffusion, i.e., thermal noise diffuses the phase of each oscillator and the mean-field orbit is progressively smeared off all over the limit cycle; ii) stable statistical mixture of two (or more) limit cycles are possible, i.e., thermal noise allows to explore more than one limit cycle, when adjacent attractors associated with the Bessel functions of Eq.~(\ref{eq:auxiliary function})~\cite{Heinrich2011,Holmes2012} are not too distant in phase space. Multistability in this case is manifested by a bimodal stationary probability distribution occupying two different limit cycles.

\begin{figure}[]
\centering
\includegraphics[width=3.5in]{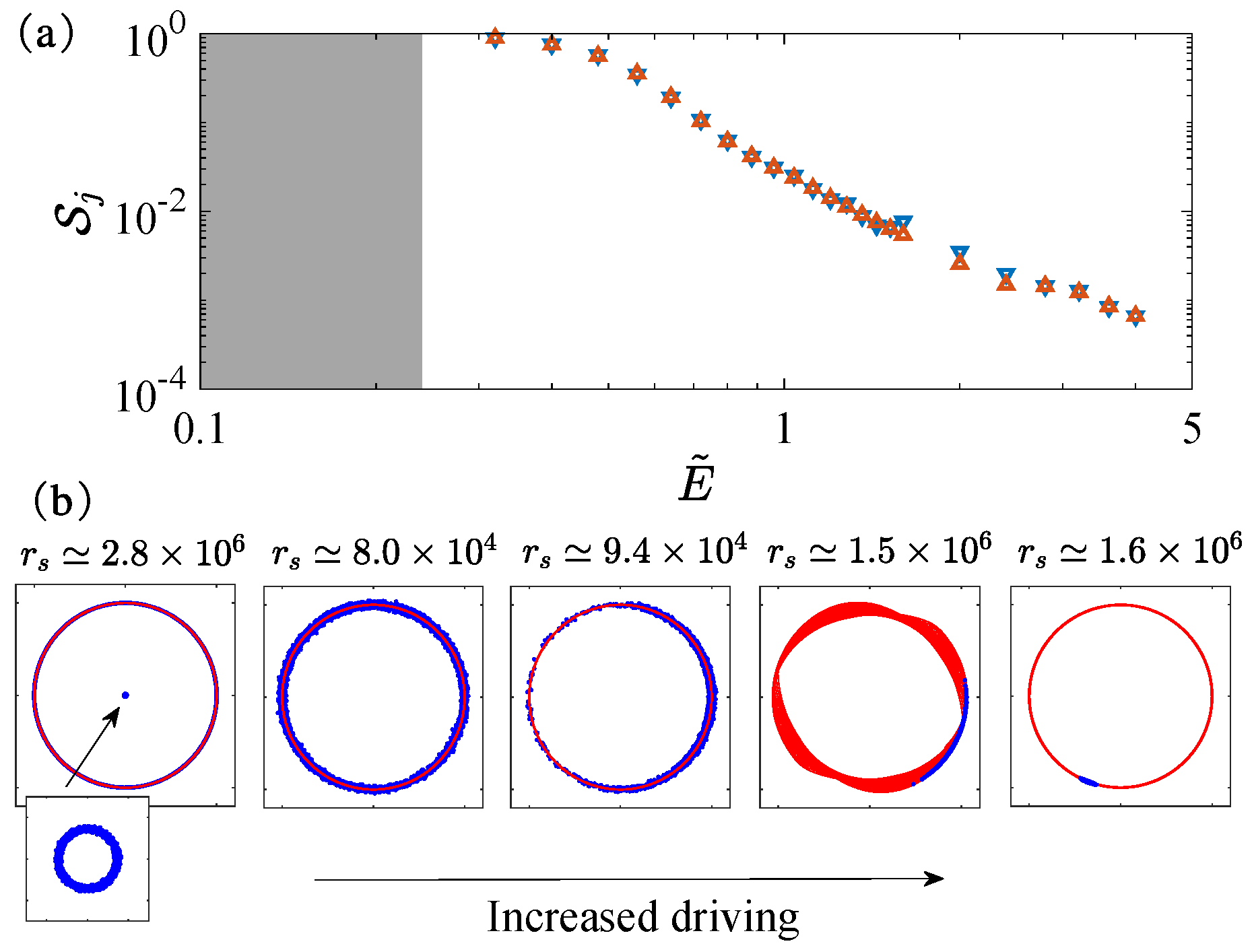}
\caption{(a): $\mathcal{S}$ of Eq.~(\protect\ref{eq: Phase fluctuation}) as a function of $\tilde{E}$ at the same time instant $t=t_1+t_2$. The results are obtained by first simulating the full classical Langevin equations up to $t_1=100/\omega_1$, and then simulating the amplitude equations up to  $t_2=2500/\Delta\omega\sim 0.5$s. The statistical results are obtained by $10000$ calculations of the stochastic equations. The grey area at small values of $\tilde{E}$ corresponds to a multi-stable region where $\mathcal{S}$ cannot be evaluated; downward blue triangles refer to oscillator $1$, while upward red triangles refer to oscillator $2$.
(b): Simulation results for the phase space probability distribution of oscillator $1$ with increased driving (from left to right we have $\tilde{E}=0.16$, $0.32$, $0.56$, $1.04$ and $4$, respectively). Blue dots denote $10000$ stochastic results at time $t$, while red lines denote limit cycle trajectory in a small time interval around $t$. In each plot $r_s$ denotes the radius of the (larger) limit cycle, and the other parameters are the same as those in Fig.~\ref{fig:2}.
\label{fig:4}}
\end{figure}
These two non-Gaussian features are shown and analysed in detail in Fig.~\ref{fig:4}, where we have fixed $\tilde{\Delta}=1$ and $\tilde{\kappa}=1$, and we consider different values of the driving amplitude $\tilde{E}$ up to $\tilde{E}=4$. Bistability occurs within a small range of values of $\tilde{E}$ (smaller than $\tilde{E}=0.24$, which corresponds to $P \simeq 0.32$ mW), and denoted by the grey area in Fig.~\ref{fig:4}(a), corresponding to a weak driving regime. A typical situation is shown in the first panel on the left in Fig.~\ref{fig:4}(b) which refers to $\tilde{E}=0.16$: one can clearly see the coexistence of two limit cycles in the phase space of one oscillator, one much smaller than the other. Moreover at the chosen time instant (about five times the mechanical relaxation time) the oscillator phase has become completely random. Therefore one has a phase invariant bimodal phase space probability distribution describing the statistical mixture of the two limit cycles. Due to the large distance in phase space, jumps between one limit cycle to the other within a given stochastic trajectory have a negligible probability.

As the drive increases, the limit cycle attractors move away from each other, and both the initial thermal distribution and thermal noise are no more able to populate simultaneously two adjacent attractors.  As a result, only one limit cycle is occupied in phase space for $\tilde{E}>0.24$, and this single ring structure remains valid up to the very large value $\tilde{E} \simeq 6$, while for (quite unrealistic) stronger drivings, multistable structures reappear. In the case of one populated limit cycle only, one can make a more quantitative analysis of phase diffusion, which is shown in Fig.~\ref{fig:4}(a) where the phase diffusion quantifier $\mathcal{S}$ of Eq.~(\ref{eq: Phase fluctuation}) for each resonator is plotted versus $\tilde{E}$.
The quantity $\mathcal{S}$ is evaluated always at the same time instant $t=t_1+t_2\sim 0.5s$ for the different values of $\tilde{E}$, and one can see a fast, monotonic decrease of phase diffusion for increasing driving, implying that phase diffusion becomes slower and slower for increasing input power. The behavior of Fig.~\ref{fig:4}(a) is well fitted by $\mathcal{S} \propto \tilde{E}^{-3} $, suggesting that diffusion time in a limit cycle $\tau_{\rm diff}$ (the time each oscillator phase takes to randomize itself over $2 \pi$) scales as $\tau_{\rm diff} \propto \tilde{E}^{3}/\bar{n}$, which is quite well reproduced by our simulations.

\begin{figure}[]
\centering
\includegraphics[width=3.5in]{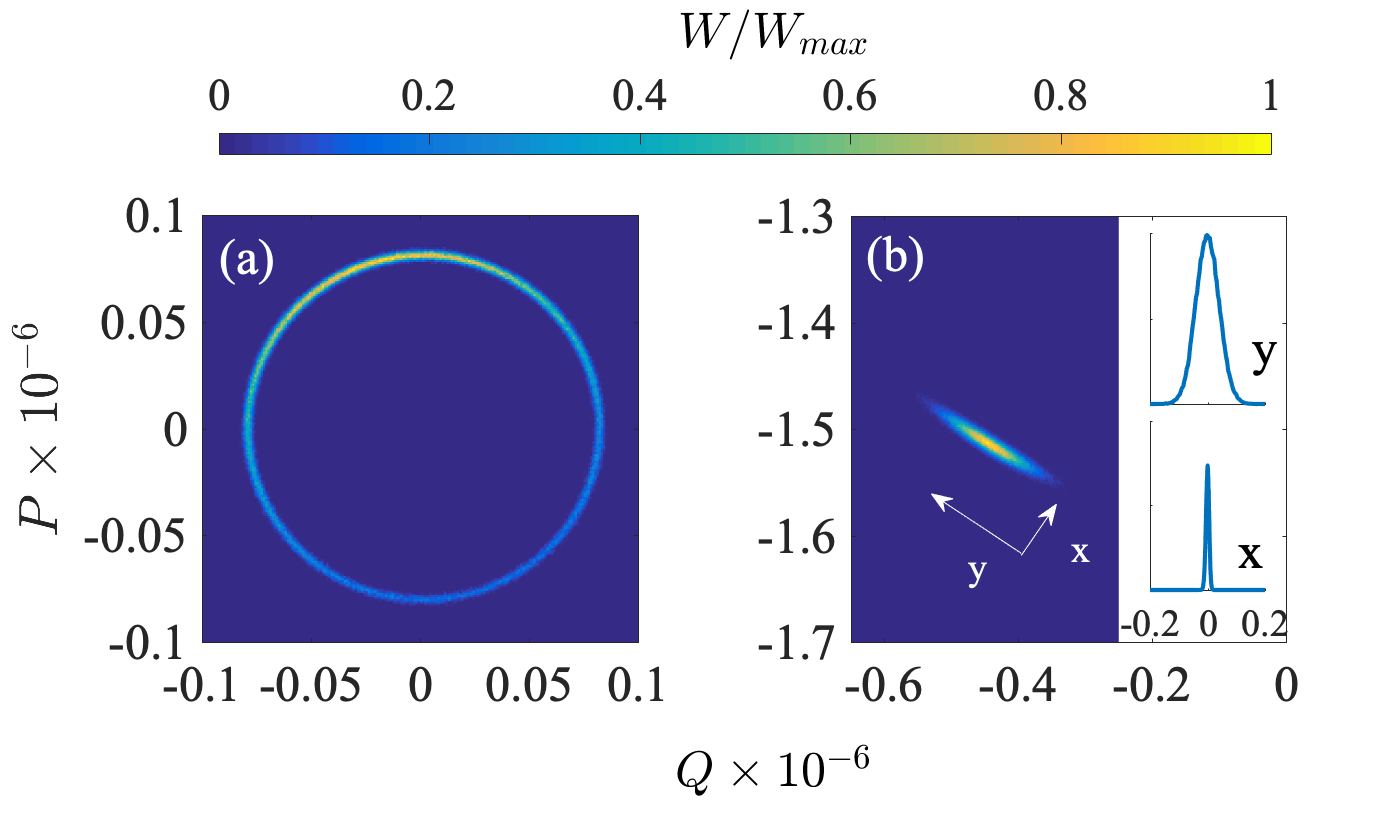}
\caption{ (a) and (b): Phase space probability distribution of oscillator $1$ with $\tilde{E}=0.32$ and $\tilde{E}=4$, respectively. (a) and (b) are obtained by $100000$ realizations of the stochastic equations. The other parameters are the same as those in Fig.~\ref{fig:4}.
\label{fig:5}}
\end{figure}

The slowing down of phase diffusion for increasing $\tilde{E}$ is visualized in a more qualitative way in Fig.~\ref{fig:4}(b). In addition to the transition from the double-ring to the single-ring structure, Fig.~\ref{fig:4}(b) shows that diffusion over the classical orbit becomes smaller and smaller with increasing pump power. Especially, when $\tilde{E}=4$, the phase variance is extremely small and one has a Gaussian-like statistics up to this time instant. This is better shown in Fig.~\ref{fig:5}(a) and (b), where we show the phase space probability distribution corresponding to the two cases where phase diffusion is significant ($\tilde{E}=0.32$), and significantly slowed down ($\tilde{E}=4$).
In particular, the corresponding probability distributions of x (amplitude) and y (phase) quadratures are also plotted in subfigure (b), and we find that they have Gaussian shapes, although with different standard deviations, meaning that for this parameter regime, a linearized Gaussian analysis is valid for quite long evolution times.

We underline however that here one can never have full suppression of phase diffusion and spontaneous symmetry breaking of time translation symmetry, as it occurs for example in the mean field analysis of synchronization of an optomechanical array of Ref.~\cite{Ludwig2013}, and that may occur only in the limit of very large number of resonators. The fact that in the long time limit full phase diffusion is achieved, and that the stationary phase space probability distribution of each resonator is always phase invariant can be also seen analytically by exploiting the bright-dark mode analysis of Sec. \ref{Bright and dark mode analysis}, at least in the simple case when $\Delta \gamma = \Delta \omega = 0$. In this case, the two mechanical resonators have equal frequencies and damping, so that $\lambda = 0$ and the bright and dark modes are uncoupled.
As a consequence, the phase space probability distribution is factorized, with the dark mode remaining in its thermal state $P_{\rm st}(A_d) \sim \exp\{-|A_d|^2/2 \bar{n}\}$, while for the bright mode one can apply the treatment for the case of a standard single-mode optomechanical system around the parametric instability~\cite{Marquardt2006,Rodrigues2010} and get
\begin{equation}\label{eqPsta}
  P_{\rm st}(A_b) \propto \exp\left[-\frac{|A_b|^2}{2 \bar{n}}\right] \exp\left[-\frac{g_q}{\gamma \bar{n}}\int_0^{|A_b|} dr \mathcal{F}_i(r) r \right].
\end{equation}
Therefore, the stationary probability distribution depends upon the moduli $|A_d|$ and $|A_b|$, and, when transforming back to the oscillator variables, only upon $I_1$, $I_2$ and $\theta_-$. The phase sum $\theta_+ = \theta_1+\theta_2$ is instead completely random, and so are the two resonator phases $\theta_1$ and $\theta_2$. When $\lambda \neq 0$ and bright and dark modes are coupled, there is no simple method for deriving the stationary phase space probability distribution, but we expect that it is still independent from $\theta_+$.

\section{Robustness with respect to thermal noise}
\label{Robustness with respect to thermal noise}
\begin{figure}[]
\centering
\includegraphics[width=3.5in]{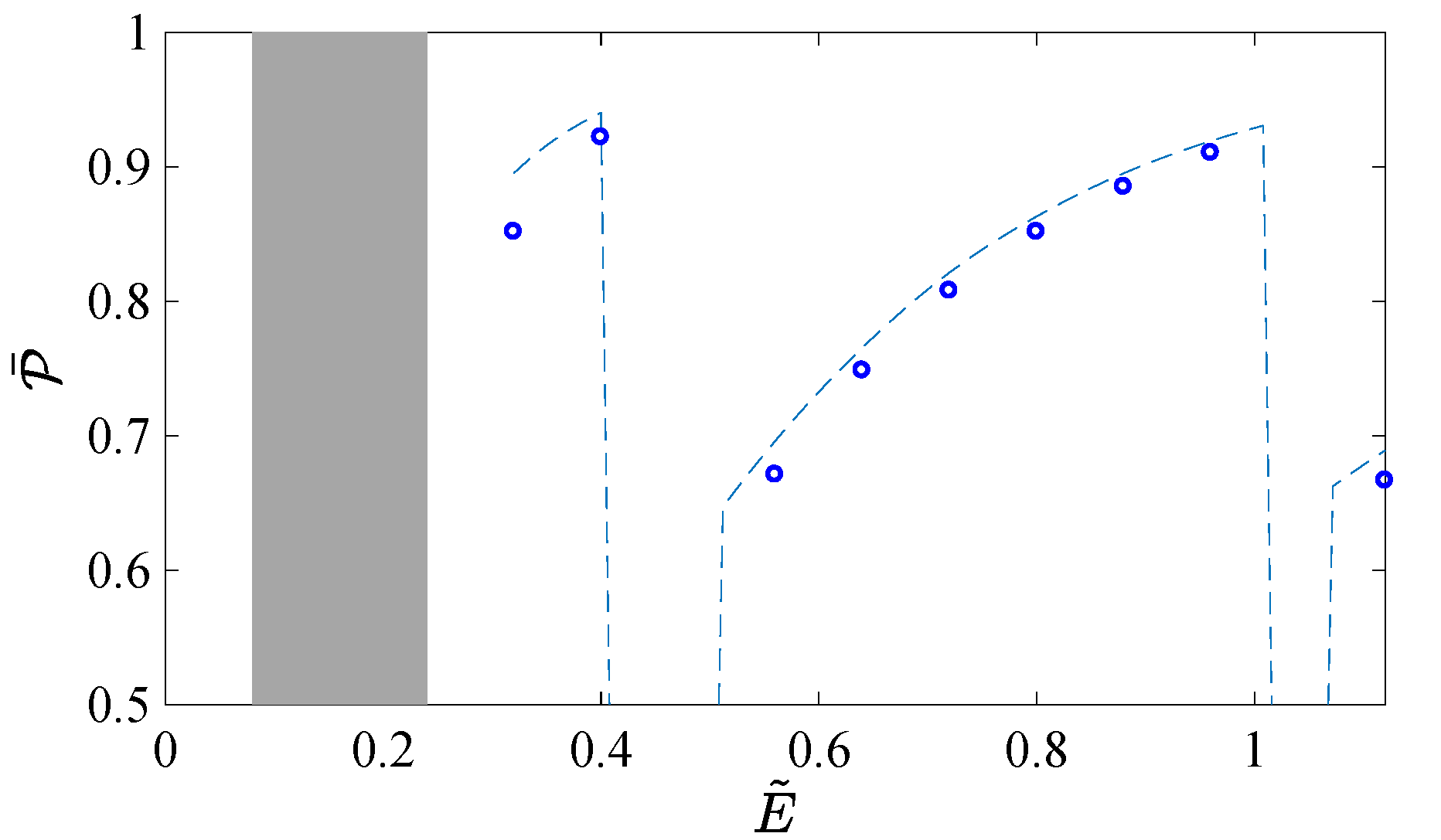}
\caption{Phase synchronization measure $\bar{\mathcal{P}}$ as a function of $\tilde{E}$, with fixed $\tilde{\Delta}=1$. Blue circles denotes the numerical result in the presence of noise and the dashed lines provides the mean-field result without noise (corresponding to those shown in Fig.~2). The value of $\bar{\mathcal{P}}$ in the presence of noise has been obtained averaging over $100$ randomly chosen trajectories within the full ensemble of $10000$ trajectories, and result did not depend upon the chosen sub-ensemble. The other parameters are the same as those in Fig.~\ref{fig:4}.
\label{fig:6}}
\end{figure}

\begin{figure}[]
\centering
\includegraphics[width=3.5in]{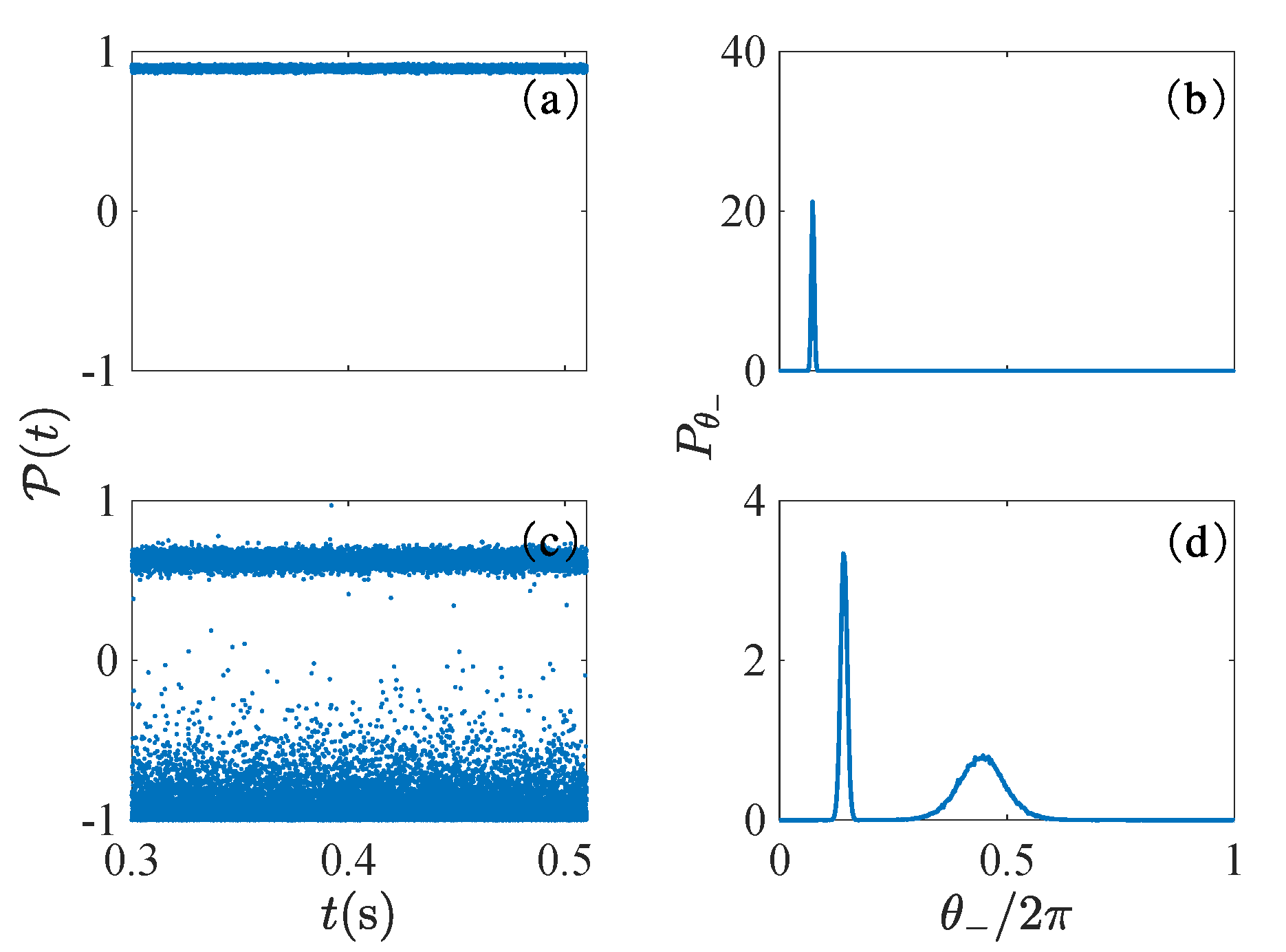}
\caption{(a) and (c): Random long time behavior of the measure $\mathcal{P}(t)$ in the case of a single limit cycle ($\tilde{E}=0.32$) and coexistence of two limit cycles ($\tilde{E}=0.16$), respectively. Each point on the trajectory is randomly selected from $10000$ results. (b) and (d) are the corresponding phase difference probability distribution obtained by considering $100000$ trajectories. The other parameters are the same as those in Fig.~\ref{fig:4}.
\label{fig:7}}
\end{figure}

We now show that, although thermal noise can significantly affect the dynamical properties of two oscillators, synchronization is very robust with respect to thermal noise. To illustrate this fact, we calculated the synchronization measure based on Eqs.~\eqref{eq:phase error measure}-(\ref{eq:m time measure}) versus the driving amplitude, and plot the results in Fig.~\ref{fig:6}. We see only a very small decrease of the synchronization measure due to noise, while the behavior remains exactly the same in the two cases, showing that synchronization in this OMS could be easily observed at room temperature and strong enough optomechanical coupling.
The robustness of synchronization with respect to thermal noise is visible also in Fig.~\ref{fig:7}, where we consider $\tilde{E}=0.32$ [(a) and (b)], corresponding to a single limit cycle, and $\tilde{E}=0.16$ [(c) and (d)], which refers to the bistable situation of Fig.~\ref{fig:4}. In Fig.~\ref{fig:7}(a) and Fig.~\ref{fig:7}(c)  we show the random long time behavior of the measure $\mathcal{P}(t)$, where each point is randomly selected from $10^4$ simulated trajectories: we see a clear robust $0$-phase synchronization in the case of a single limit cycle. In Fig.~\ref{fig:7}(c) we have a statistical mixture of two limit cycles, one $0$-phase synchronized, $\theta_-=0$, and one with $\theta_-\sim \pi$, and larger fluctuations of the measure $\mathcal{P}(t)$. In Fig.~\ref{fig:7}(b) and \ref{fig:7}(d), we plot the corresponding probability distribution of the phase difference at time $t\simeq 0.5$, and we see that synchronization is robust, especially for $0$-phase synchronized limit cycles, because in these cases $P_{\theta}$ is extremely peaked, with a very small uncertainty. In this respect, the $\pi$-synchronized limit cycle in the bistable case is much less robust. The different values of the average relative phase and therefore the different kind of synchronization in the bistable case is not surprising because, as we have seen in Sec.~V, this value strongly depends upon the values of the two limit cycle amplitudes, $I_1$ and $I_2$, which are very different in the two cases.

\begin{figure}[]
\centering
\includegraphics[width=3.5in]{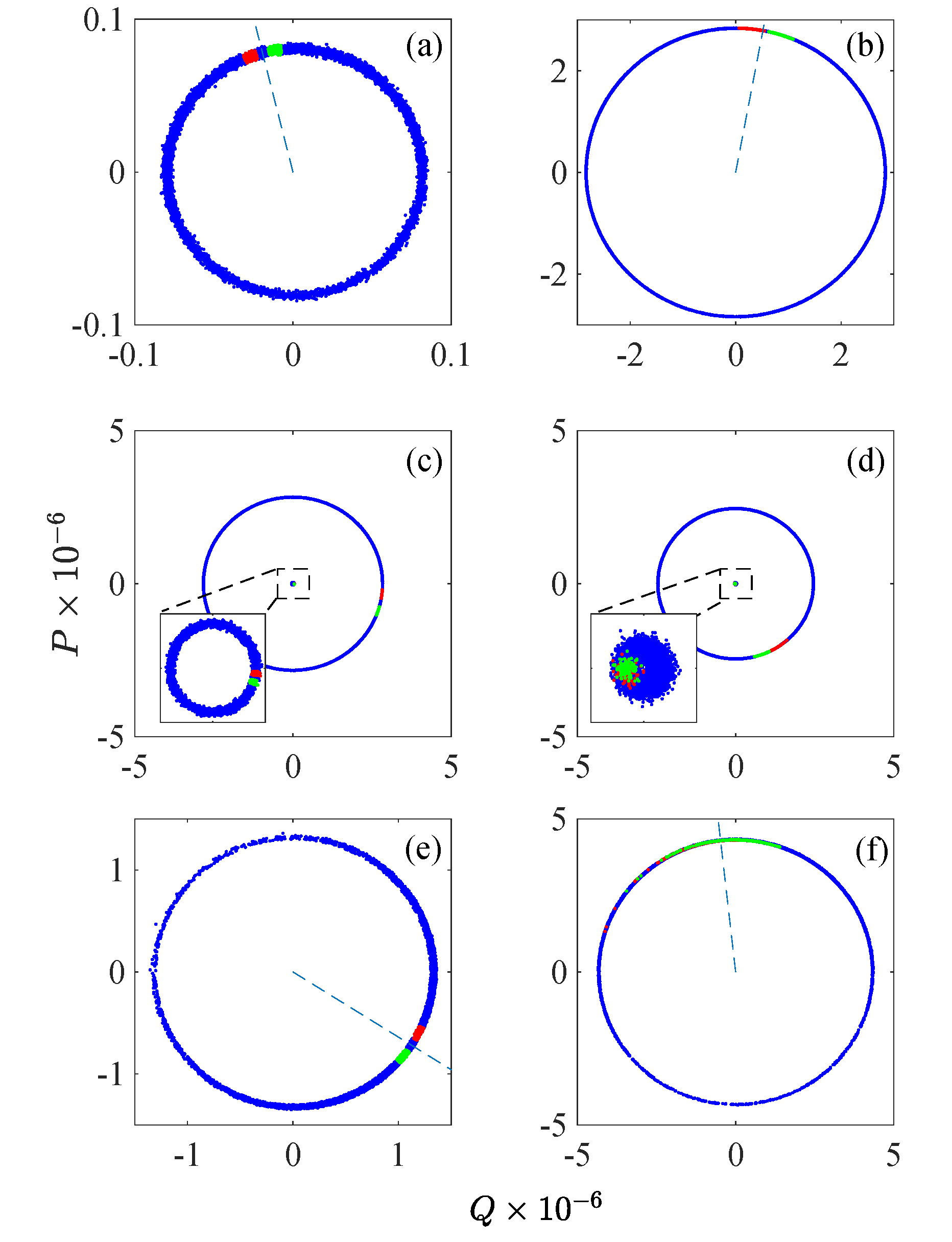}
\caption{Phase space probability distribution of oscillator $1$ (left column) and $2$ (right column) corresponding to single limit cycle synchronization ($\tilde{E}=0.32$, (a) and (b)), statistical mixture of two synchronized limit cycles ($\tilde{E}=0.16$, (c) and (d)), and no synchronization ($\tilde{E}=0.48$, (e) and (f)). The blue dots represent the reduced probability distribution at a given time after a large number of trajectories, the blue dashed lines in (a,b,e,f) denote the mean value of the oscillator phase, i.e., $\phi_j=\arg[\langle A^i_j \rangle ]$. The green (red) narrow phase interval within which we select a sub-ensemble of points of oscillator $1$ in the left column is defined as $\phi_1 + 0.05 \leq \theta_1 \leq \phi_1+0.15$ ($\phi_1 - 0.15 \leq \theta_1 \leq \phi_1-0.05$). In the right column the green (red) points denote the corresponding conditional value of the oscillator phase $\theta_2$. The other parameters are the same as those in Fig.~\ref{fig:3}.
\label{fig:8}}
\end{figure}
The above analysis and especially Fig.~\ref{fig:7} shows that, even though the phase of each oscillator tends to be diffused all over $2\pi$ at long times, in the presence of synchronization the two phases are strongly locked to each other, with very small fluctuations of the relative phase $\theta_-$ around a fixed value, even in the presence of large thermal noise.
In Fig.~\ref{fig:8} we illustrate in more detail this phase locking by looking at conditional phase space probability distributions. We plot the reduced Wigner functions in
phase-space at a large time $t$ for oscillator $1$ (left column) and oscillator $2$ (right column) for three different values of the driving amplitude, $\tilde{E} = 0.32$
[Fig.~\ref{fig:8}(a)-(b)], $\tilde{E} = 0.16$ [Fig.~\ref{fig:8}(c)-(d)], and $\tilde{E} = 0.48$ [Fig.~\ref{fig:8}(e)-(f)].
The blue dots represent the phase-space probability distribution after a large number of trajectories [for example Fig.~\ref{fig:8}(a) coincides with Fig.~\ref{fig:5}(a)] and they all show full phase diffusion over the limit cycle of each oscillator since we are in the regime of not too large $\tilde{E}$. However, if we select a sub-ensemble of phase space points for oscillator $1$ in a narrow interval of its phase $\theta_1$, we see that, at least for Figs.~\ref{fig:8}(a)-(d),  the corresponding phase space points for oscillator $2$ also lie within a narrow interval of $\theta_2$. This is consistent with the presence of synchronization and with the results of Fig.~\ref{fig:7}, because, due to the locked value of $\theta_-$, if we fix $\theta_1$, also the other oscillator phase is determined with high probability. In particular, the green (red) points on the left column denote the chosen narrow interval for oscillator $1$, with a clockwise (counter-clockwise) deviation with respect to the average phase, and the points with the same colour in the plot in the right column denote the corresponding conditioned points for oscillator $2$. The two intervals are very narrow for both oscillators, clearly showing the strong phase correlations in Figs.~\ref{fig:8}(a)-(d), where synchronization occurs. This occurs both for $\tilde{E} = 0.16$ and $\tilde{E} = 0.32$, i.e., either in the monostable and bistable case, even if, as already suggested by Fig.~\ref{fig:7}(d), phase correlation is weaker in the case of the $\pi$-phase synchronized resonators, and the conditional $\theta_2$ interval is less narrow. We also notice that the center of green (red) intervals in sub-figures (b) and (d), regardless of $0$-synchronization or $\pi$-synchronization, always have clockwise (counter-clockwise) deviations relative to the average phase. In other words, anti-phase locking does not occur in this case.

Finally, it is worth focusing on Figs.~\ref{fig:8}(e)-(f), which refer to $\tilde{E}=0.48$ and, as shown also in Fig.~\ref{fig:6}, corresponds to \emph{unsynchronized} resonators. What is remarkable here is that, despite the absence of synchronization, from the conditional green and red dots we already see a clear phase correlation between the two resonators, even if weaker than the one manifested in the figures above corresponding to a nonzero synchronization measure. Therefore a weak form of phase locking acts as a sort of precursor of synchronization; when $\tilde{E}=0.48$, one has that the phase difference $\theta_-$ tends to assume a definite value, with a small variance, but it does not assume a time-independent value yet, so that the two oscillators do not become synchronized.

\section{Conclusions}
We have explored the effects of noise on the synchronization of an OMS formed by two mechanical resonators coupled to the same driven optical cavity mode. The dynamics has been studied by first adopting classical Langevin equations for the optical and mechanical complex amplitudes, from which we have then derived the corresponding stochastic equations for the slowly varying amplitude equations for the mechanical resonators only. These latter equations can be used to study the effective synchronization dynamics in the long time limit and neglecting transient regimes. We have introduced effective bright and dark mechanical amplitudes, which allow us to simplify the physical description of the system dynamics.

We have first studied the rich synchronization phase diagram in the noiseless case, and then the effects of thermal noise on such a diagram. We have studied phase diffusion of the two oscillator phases in which the limit cycle of each oscillator is progressively smeared off by thermal noise.
We have seen that phase diffusion is significantly slowed down by increased driving, and that for large enough driving a Gaussian linearized treatment is valid for intermediate times, even though phase diffusion is never fully suppressed and we did not expect any spontaneous symmetry breaking of time translation symmetry.  A second non-Gaussian feature due to thermal noise and for weak driving is the presence of a statistical mixture of two different coexisting synchronized limit cycles, with different amplitudes and relative phase.

In general we find that synchronization in the present OMS is very robust to thermal noise: the adopted synchronization measure shows only a very small decrease due to the presence of noise, and the synchronization phase diagram remains practically unaffected. Therefore synchronization of the two mechanical resonators should be visible at room temperature and not too small optomechanical cooperativities. In the presence of synchronization the two oscillator phases are locked to each other, and this correlation is weakly affected by thermal noise, as we illustrate also by means of oscillators' conditional phase space distributions. Interestingly, we find that phase locking may occur also when the two oscillators are not synchronized, suggesting that an emerging nonzero phase correlations between the two resonators may be considered as a precondition for synchronization.

\begin{acknowledgements}
We acknowledge the support of the European Union Horizon 2020 Programme for Research and Innovation through the Project No. 732894 (FET Proactive HOT) and the Project QuaSeRT funded by the QuantERA ERA-NET Cofund in Quantum Technologies. P. Piergentili acknowledges support from the European Union's Horizon 2020 Programme for Research and Innovation under grant agreement No. 722923 (Marie Curie ETN - OMT).
\end{acknowledgements}

\end{document}